\documentclass[useAMS,usenatbib]{mn2e}

\usepackage{amssymb}
\usepackage[fleqn]{amsmath}
\usepackage{times}
\usepackage{graphicx}
\usepackage{multirow}
\usepackage{rotating}
\usepackage{bm}

\newcommand{\smo}{Smol\v{c}i\'{c}}

\def\f#1   {Fig.~\ref{#1}}
\def\s#1   {Sec.~\ref{#1}}
\def\tab#1   {Tab.~\ref{#1}}
\def\eq#1   {Eq.~\ref{#1}}
\def\t#1   {Tab.~\ref{#1}}

\def\comm#1   {{\tt (COMMENT: #1) }}

\def\chandra{{\it Chandra}}
\title[New insights from deep VLA data on CID-42]{New insights from deep VLA data on the potentially recoiling black hole CID-42 in the COSMOS field}

\author[Novak \ et al. ]{Mladen Novak$^{1}$\thanks{E-mail:mlnovak@phy.hr},
Vernesa Smol{\v c}i{\'c}$^{1}$,
Francesca Civano$^{2,3}$,
Marco Bondi$^{4}$,
Paolo Ciliegi$^{5}$,
\newauthor 
Xiawei Wang$^{6}$,
Abraham Loeb$^{6}$,
Julie Banfield$^{7,8}$,
Stephen Bourke$^{9}$,
Martin Elvis$^{10}$,
\newauthor 
Gregg Hallinan$^{9}$,
Huib T. Intema$^{11}$,
Hans-Rainer Kl\"ockner$^{12,13}$,
Kunal Mooley$^{9}$,
\newauthor 
Felipe Navarrete$^{14}$
\\
$^{1}$University of Zagreb, Physics Department, Bijeni\v{c}ka cesta 32, 10002 Zagreb, Croatia\\
$^{2}$Yale Center for Astronomy and Astrophysics, 260 Whitney ave, New Haven, CT 06520, USA\\
$^{3}$Smithsonian Astrophysical Observatory, 60 Garden st., Cambridge, MA 02138, USA\\
$^{4}$INAF - Istituto di Radioastronomia di Bologna, via P. Gobetti, 101, I-40129, Bologna, Italy\\
$^{5}$INAF - Osservatorio Astronomico di Bologna, Via Ranzani 1, I-40127 Bologna, Italy\\
$^{6}$Department of Astronomy, Harvard University, 60 Garden St., Cambridge, MA 02138, USA\\
$^{7}$CSIRO Australia Telescope National Facility, PO Box 76, Epping, NSW, 1710, Australia\\
$^{8}$Research School of Astronomy and Astrophysics, Australian National University, Weston Creek, ACT 2611, Australia\\
$^{9}$Department of Astronomy, California Institute of Technology, MC 249-17, 1200 East California Blvd, Pasadena, CA 91125, USA\\
$^{10}$Harvard-Smithsonian Center for Astrophysics, 60 Garden Street, Cambridge, MA 02138, USA\\
$^{11}$National Radio Astronomy Observatory, 1003 Lopezville Road, Socorro, NM 87801-0387, USA\\
$^{12}$Subdepartment of Astrophysics, University of Oxford, Denys-Wilkinson Building, Keble Road, Oxford OX1 3RH, UK\\
$^{13}$Max-Planck-Institut f\"{u}r Radioastronomie, Auf dem H\"{u}gel 69, D-53121 Bonn, Germany\\
$^{14}$Argelander Institute for Astronomy, Auf dem H\"{u}gel 71, Bonn, 53121, Germany\\
}

\begin{document}

% \date{Accepted 20XX Month XX. Received 20XX Month XX; in original form 20XX Month XX}

\pagerange{\pageref{firstpage}--\pageref{lastpage}} \pubyear{2014}

\maketitle

\label{firstpage}

\begin{abstract}
We present deep 3~GHz VLA observations of the potentially recoiling black hole CID-42 in the COSMOS field.
This galaxy shows two optical nuclei in the HST/ACS image and a large velocity offset of $\approx$ 1300 km s$^{-1}$ between the broad and narrow H$\beta$ emission line although the spectrum is not spacially resolved (\citealt{civano10}).
The new 3~GHz VLA data has a bandwidth of 2~GHz and to correctly interpret the flux densities imaging was done with two different methods: multi-scale multi-frequency synthesis and spectral windows stacking.
The final resolutions and sensitivities of these maps are 0.7$\arcsec$ with rms = $4.6\,\mu$Jy\,beam$^{-1}$ and 0.9$\arcsec$ with rms = $4.8\,\mu$Jy\,beam$^{-1}$ respectively.
With a 7$\sigma$ detection we find that the entire observed 3~GHz radio emission can be associated with the South-Eastern component of CID-42, coincident with the detected X-ray emission.
We use our 3~GHz data combined with other radio data from the literature ranging from 320~MHz to 9~GHz, which include the VLA, VLBA and GMRT data, to construct a radio synchrotron spectrum of CID-42. The radio spectrum suggests a type~I unobscured radio-quiet flat-spectrum AGN in the South-Eastern component which may be surrounded by a more extended region of old synchrotron electron population or shocks generated by the outflow from the supermassive black hole.
Our data are consistent with the recoiling black hole picture but cannot rule out the presence of an obscured and radio-quiet SMBH in the North-Western component.

\end{abstract}

\begin{keywords}
galaxies: nuclei - galaxies: active - galaxies: interactions - galaxies: individual: CID-42 - radio continuum: galaxies
\end{keywords}

\section{Introduction}
\label{sec:intro}
During galaxy major mergers, the central supermassive black holes (SMBHs) that reside within the merging galaxies will 
form a bound binary SMBH system that can further merge (e.g. \citealt{volonteri03}, \citealt{hopkins08}, \citealt{colpi09}). 
At the time of the SMBH binary coalescence, strong gravitational wave (GW) radiation is emitted an-isotropically, 
depending on the spin and mass-ratio of the two SMBHs, and to conserve linear momentum, the newly formed single SMBH recoils (\citealt{peres62}, \citealt{bekenstein73}).
Recoiling SMBHs are the direct products of processes in the strong field regime of gravity and they are one of the key observable signatures of a SMBH binary merger.  As the SMBH recoils from the center of the galaxy, the closest regions (disk and broad line regions) are carried with it and the more distant regions are left behind depending on the recoil velocity (\citealp{merritt06}, \citealp{loeb07}). Because GW recoil displaces (or ejects) SMBHs from the centers of galaxies, these events have the potential to influence the observed 
co-evolution of SMBHs with their host galaxies, as demonstrated by numerical simulations (\citealt{blecha11}, \citealt{sijacki11}, \citealt{guedes11}).

Only few serendipitous discoveries of recoiling candidates have been reported in the literature (\citealt{komossa08}, \citealt{shields09}, \citealt{robinson10}, \citealt{jonker10}, \citealt{batcheldor10}, \citealt{steinhardt12}, \citealt{bianchi13}, \citealt{koss14}) and systematic observational searches have  resulted in no candidates so far (\citealt{bonning07}, \citealt{eracleous12}, \citealt{komossa12}). 

The \chandra-COSMOS source \allowbreak{CXOC J100043.1+020637} ($z=0.359$, \citealt{elvis09}, \citealt{civano12}), also known as CID-42, is a candidate for being a GW recoiling SMBH with both imaging (in optical and X-ray) and spectroscopic recoil signatures (\citealt{civano10, civano12b, civano12}). 
CID-42 shows two components separated by $\approx 0.5^{\prime\prime}$ ($\approx 2.5$ kpc; see below for cosmology details\footnote{Conversions between arcseconds and kiloparsecs are done according to \cite{wright2006} using the assumed cosmology.}) in the Hubble Space Telescope Advanced Camera for Surveys
(HST/ACS) image and embedded in the same galaxy.
As presented in \cite{civano10} and \cite{civano12b}, the South-Eeastern (SE) optical source has a point-like morphology typical of a 
bright active galactic nucleus (AGN) and it is responsible for the entire ($>$97\%) X-ray emission 
of this system. The Noth-Western (NW) optical source has a
more extended profile in the optical band with a scale length of $\approx 0.5$\,kpc, and the upper limit measured for its 
X-ray emission is consistent with being produced by star-formation.
In the optical spectra of CID-42 (VLT, Magellan, SDSS and DEIMOS; see \citealt{civano12b, civano12}), a velocity offset of $\approx 1300$ km\,s$^{-1}$ is
measured between the broad and narrow components of the H$\beta$ line (Figure 5 and 6 of \citealt{civano10}).

Despite diverse scenarios being proposed to explain the nature of CID-42 (e.g., \citealt{comerford09}, \citealt{civano10}),  
the upper limit on the X-ray luminosity combined with the analysis of the multiwavelength spectral energy distribution favors 
the GW recoil scenario, although the presence of a very obscured SMBH in the NW component cannot be fully ruled out.
The current data are consistent with a recoiling SMBH ejected approximately 1 -- 6 Myr ago, as shown 
by detailed modeling presented in \cite{blecha13}. 

In this paper, we present new deep data at 3~GHz from the Karl G. Jansky Very Large Array (VLA) of the National Radio Astronomy Observatory (NRAO). For the analysis we also make use of the radio data from the literature (\citealt{schinnerer07}, \citealt{wrobel14}, \citealt{smolcic2014}, Karim et al., in prep.) to study the radio emission in CID-42 and bring further constraints on the nature of this source.
Throughout the paper, a WMAP seven-year cosmology \citep{spergel2007,larson2011} with $H_0 =71$\,km\,s$^{-1}$\,Mpc$^{-1}$, $\Omega_M$ = 0.27, and $\Omega_\Lambda$ = 0.73 is assumed. 

\section{VLA 3~GHz data}
\subsection{Observations and reduction}
\label{sec:obs}

We make use of the observations of the first 130 hours of the Cosmic Evolution Survey (COSMOS; \citealt{scoville07a}) field with the VLA in S-band (3~GHz, 10 cm; VLA-COSMOS 3~GHz Large Project, {\smo}  et al., in prep.).
 The 2~GHz bandwidth is comprised of 16~$\times$~128 MHz spectral windows (SPWs).
We observed the 2 sq. deg. field with 110 hours in  A-configuration and 20 hours in C-configuration between 2012 and 2013.  In order to cover the entire field, we required a total of 64 pointings. 
The quasar J1331+3030 was used for flux and bandpass calibration and  J1024-0052 for gain and phase calibration. Weather conditions were excellent during the A-array observations while the C-array occasionally suffered from Summer thunderstorms.

Calibration was done using the AIPSLite pipeline (e.g. \citealt{bourke2014proc}) originally developed for the VLA Stripe 82 survey (Mooley et al. 2014, in prep). The calibrated data was then exported to CASA\footnote{http://casa.nrao.edu} (version 4.2.1), where flagging (clipping in amplitude) was done separately for the A- and C-configuration data. All the epochs were concatenated into a single measurement set used for the imaging stage described below.

\subsection{Imaging}
\label{sec:imaging}
To study the radio properties of CID-42  at 3~GHz, we imaged eight pointings (P19, P22, P30, P31, P36, P38, P39, P44) individually and joined them together in a mosaic (see  {\smo}  et al. in prep. for more details). CID-42 is located near the center of pointing P36.
Due to the wide bandwidth of the data, imaging can be done in various ways. One approach involves dividing the entire bandwidth into narrower SPWs which are then imaged separately and finally stacked together in the image plane (for details see \citealt{condon12}). This method gives accurate flux densities in each SPW because the flux density is approximately constant inside a  sufficiently small bandwidth. 
Downsides include i) loss of resolution as the lowest frequency SPW determines the resolution for the final stack, and ii) uncleaned sources which do not have a sufficiently large signal-to-noise ratio inside a single SPW.
We describe the application of this method in Section \ref{sec:spws}.

Another imaging approach is the multi-scale multi-frequency (MSMF) method developed by \cite{rau11} which allows the usage of the entire bandwidth at once to calculate the monochromatic flux density at a chosen frequency. It uses Taylor term expansion in frequencies along with multiple spatial scales to deconvolve the map and includes a map of spectral indices calculated from the Taylor series. Throughout the paper we define spectral index $\alpha$ as $F_\nu \propto \nu^\alpha$, where $F_\nu$ is flux density at frequency $\nu$. The final resolution is not limited to the lowest frequency beam size because all of the SPWs are used in the deconvolution. 
For more details on this algorithm see \citet{rau11} and \citet{rau14}.
 We describe this approach in Section \ref{sec:msmf}. We applied both imaging methods using the CASA task {\tt CLEAN}. 
\subsection{Spectral windows stack}
\label{sec:spws}

To correctly stack maps, it is necessary that they all have the same resolution which is defined by the lowest frequency spectral window (SPW). We used the robust parameter in the Briggs weighting scheme to achieve similar resolutions in different SPWs. With higher robust value (towards natural weighting) it is generally expected to get lower rms noise and a larger synthesized beam, but worse sidelobes. However, sidelobe contamination started to drastically degrade the map quality after a certain robust value. 
The final robust values we converged on are listed in Table \ref{tab:clean}. They give the optimum tradeoff between resolution, rms and sidelobes. After each SPW is cleaned, it is convolved to a common resolution of 0.9$\arcsec$ using the CASA toolkit function { \tt convolve2d} and then corrected for the primary beam (PB) response with the CASA task {\tt impbcor}.

\begin{table}
	\caption{Summary of robust parameters, flagged data and final rms values for each SPW. The rms noise is calculated after convolution to a common beam size but before primary beam correction. Last column shows which array data was entirely flagged due to RFI. }
	 \label{tab:clean}
	 
\begin{center}
	\begin{tabular}{|c|c|c|c|c|}
	\hline
	SPW & $\nu $ (GHz) & Robust & RMS ($\mu$Jy\,beam$^{-1}$) & Flagged \\
	\hline
	0 & 2.05   & 0.0    &   25.6  &  -     \\
	1 & 	-  & -      &    -    	   & A+C   \\
	2 & 	-  & -      &    -    	   & A+C   \\
	3 & 2.44   & 0.0    &   20.2  &  -     \\
	4 & 2.56   & 0.5    &   16.5  &  -     \\
	5 & 2.69   & 0.5    &   17.4  &  -     \\
	6 & 2.82   & 0.5    &   17.1  &  -     \\
	7 & 2.95   & 0.5    &   16.0  &  -     \\
	8 & 3.05   & 0.5    &   16.0  &  -     \\
	9 & 3.18   & 0.5    &   14.4  &  -     \\
	10& 3.31   & 1.0    &   13.0  &  -     \\
	11& 3.44   & 1.0    &   12.8  &  -    \\
	12& 3.56   & 1.0    &   12.8  &  -     \\
	13& 3.69   & 1.5    &   19.5  & C     \\
	14& 3.82   & 1.5    &   30.9  & C     \\
	15& 3.95   & 1.5    &   32.0  & C     \\
\hline
	\end{tabular} 
\end{center}

\end{table}

To optimize the S/N in the final stack, the SPW maps have been stacked  weighting each pixel by the inverse of its local noise squared.
Thus, each pixel in the sum is assigned a weight of $w=(A(\rho)/\sigma)^2$, where $A(\rho)$ is the primary beam response, $\rho$ is the distance from the pointing center and $\sigma$ is the rms.
Note that the primary beam response entering this term only scales the rms ($\sigma$ drawn from the cleaned map prior to primary beam correction) as a function of distance from the pointing center.
Because the FWHM of the primary beam (and therefore noise) changes between SPWs, not all frequencies will contribute the same amount to each pixel in the final stack. To account for this effect, a map of effective frequencies is also created by averaging the SPW frequencies using corresponding noise weights for each SPW with the CASA task {\tt immath}.
In total we produced three stacks (at a resolution of 0.9$\arcsec$): one for the full bandwidth using all of the SPWs (SPWs 0 to 15) reaching an  rms of 4.8 $\mu$Jy\,beam$^{-1}$, one for the low frequency sideband (SPWs 0 to 8) reaching an rms of 6.9 $\mu$Jy\,beam$^{-1}$, and one for the high frequency sideband (SPWs 9 to 15) reaching an  rms of 6.5 $\mu$Jy\,beam$^{-1}$. In Figure \ref{fig:cutout} we show cutouts centered on CID-42 in each map we created. Resolution, rms and effective frequency of these maps are listed in Table \ref{tab:radio}.

\subsection{Multi-scale multi-frequency}
\label{sec:msmf}

\begin{figure}
\centering
\includegraphics[width=0.35\columnwidth]{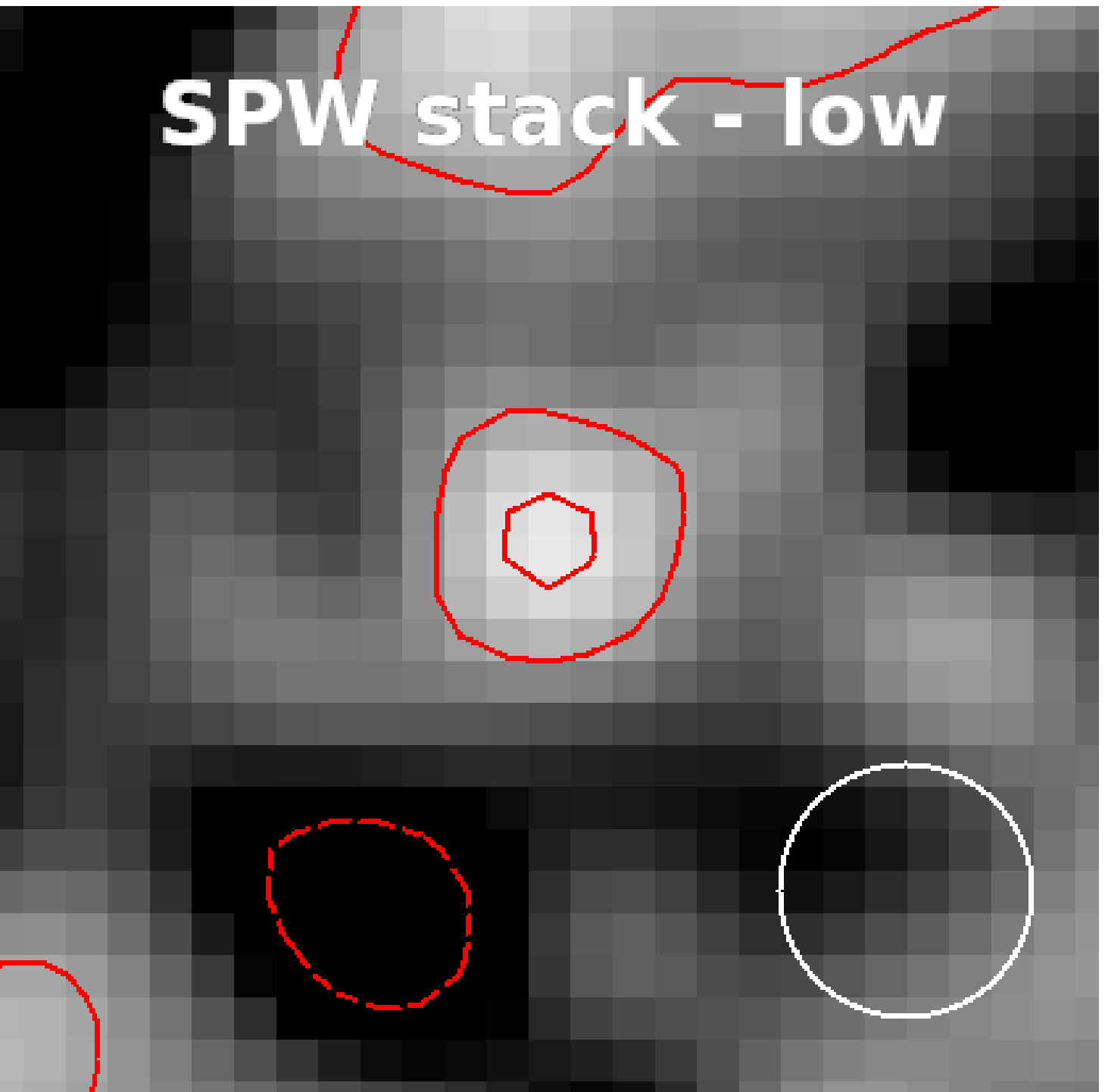}
\hspace{5pt}
\includegraphics[width=0.35\columnwidth]{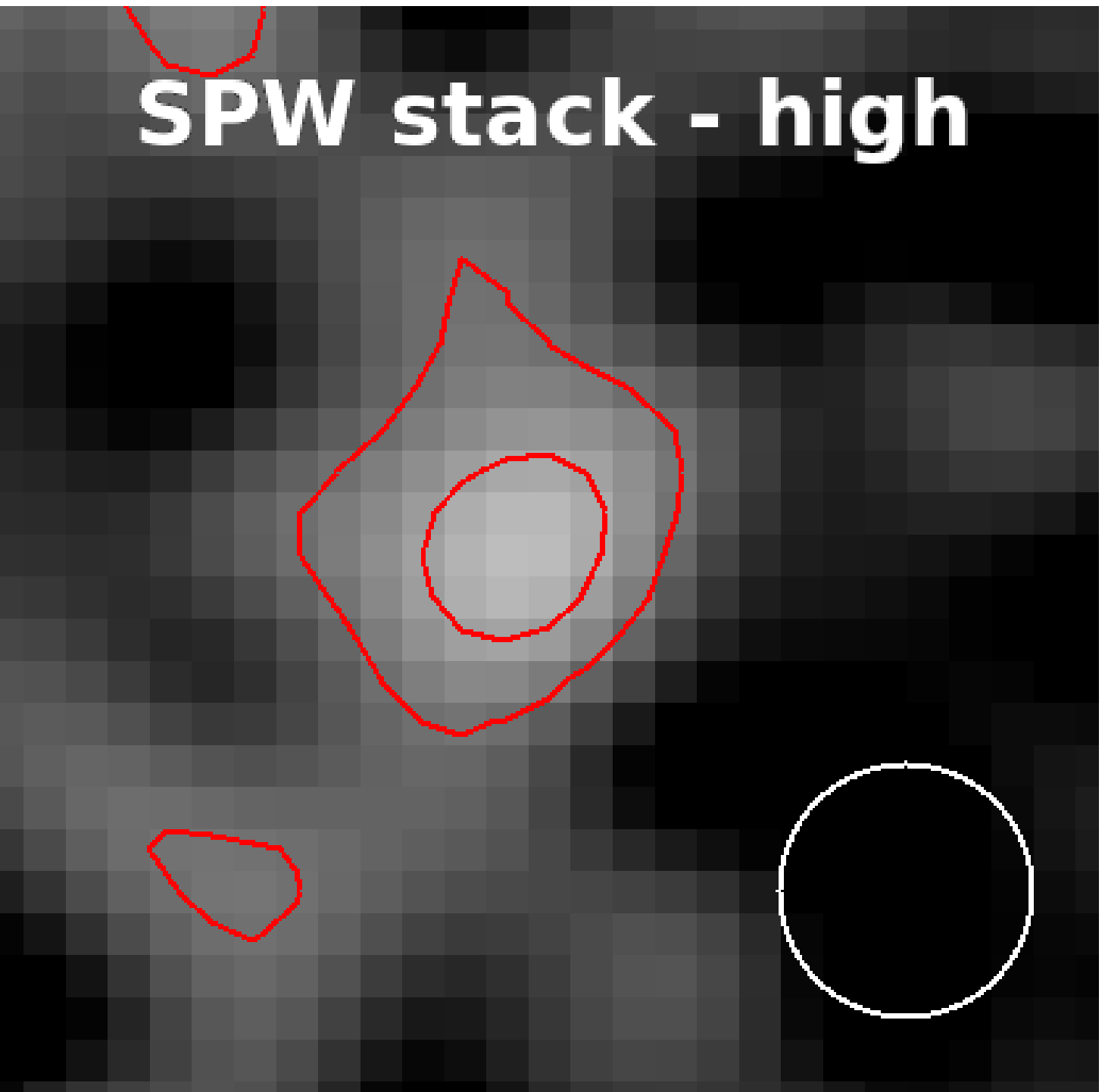}
\vspace{5pt}

\includegraphics[width=0.35\columnwidth]{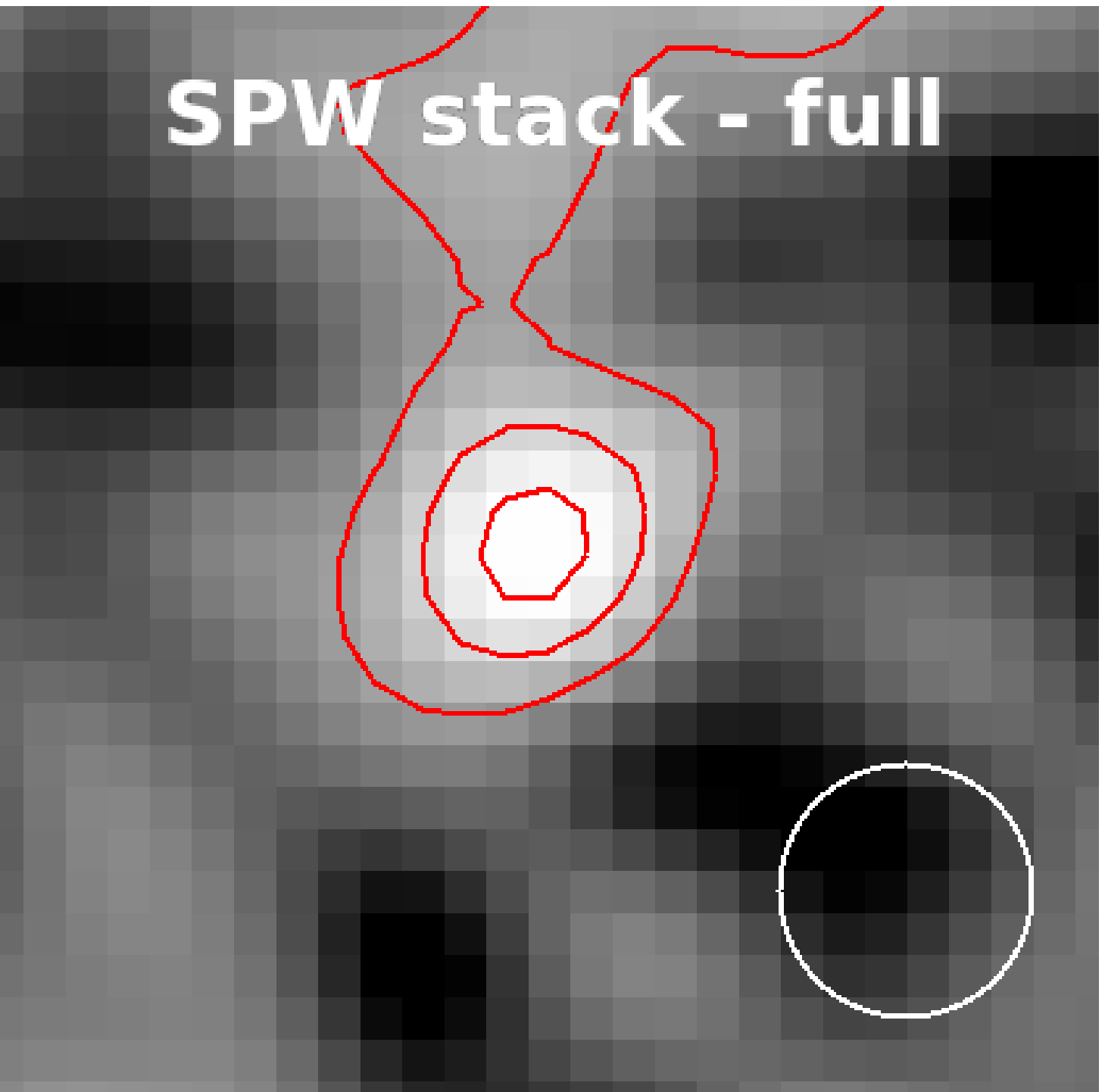}
\hspace{5pt}
\includegraphics[width=0.35\columnwidth]{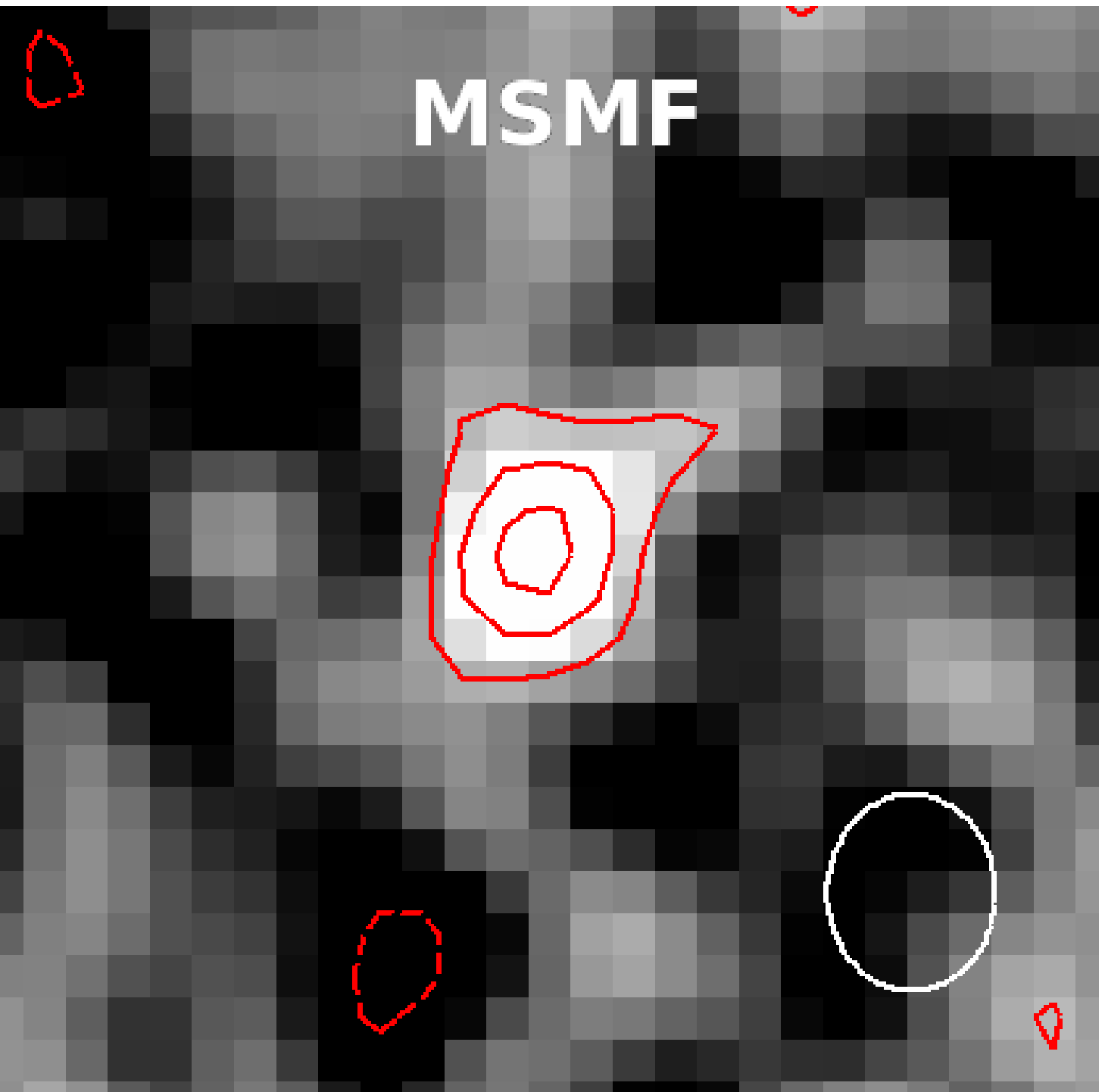}
\caption{Cutouts of 4$\arcsec$ on the side centered on CID-42: low frequency SPW stack, high frequency SPW stack, full SPW stack and MSMF map. All SPW stacks have the same resolution of 0.9$\arcsec$, while the MSMF map has a resolution of 0.7$\arcsec$. The synthesized beam size is shown in the lower right corner of each panel.
Contours of $\pm 2\sigma$, $\pm 4\sigma$ and $\pm 6\sigma$ are overplotted in red (negative contours are drawn with dashed lines). Local rms in each map is listed in Table \ref{tab:radio}.}
\label{fig:cutout}
\end{figure}

Imaging was independently performed with MSMF using the CASA task {\tt clean} that uses the entire 2~GHz bandwidth at once \citep{rau11}. Each pointing was cleaned individually. Two Taylor terms were used in the frequency expansion ({\tt nterms=2}) along with three resolution scales. The final synthesised beam size was $0.7\arcsec \times 0.6\arcsec$. We used Briggs weighting with a robust value of 0.5 to produce a map with low sidelobe contamination and good rms. After the deconvolution, wideband primary beam correction was applied using the CASA task {\tt widebandpbcor}. The resulting maps include flux densities at a reference frequency of 3~GHz along with the spectral index map ($\alpha$ map) all corrected for the PB response. The rms of the MSMF map is 4.6 $\mu$Jy\,beam$^{-1}$ and a MFMS map stamp centered at CID-42 is shown in Figure \ref{fig:cutout}.

\subsection{CID-42: Flux and size at 3~GHz}
\label{sec:sizes}
To properly extract the flux density of CID-42, information about its size is needed.  A source is considered resolved if it is larger than the beam size, however, this is a function of signal-to-noise ratio as described in \cite{bondi03}.
Thus, to estimate whether CID-42 is resolved, we need to know how the integrated-to-peak flux density ratio $S_{\mbox{\scriptsize int}}/S_{\mbox{\scriptsize peak}}$ behaves as a function of S/N in our mosaic.
For this purpose we use the AIPS task {\tt SAD} (search and destroy) to extract a catalog of sources, their positions and flux densities over the mosaic.
The catalog extraction was performed in the same way as for the other VLA-COSMOS surveys, described in detail in \citet{schinnerer07, schinnerer10} and \cite{smolcic2014}.
We have compared the peak flux density with the integrated flux density of CID-42, as well as its peak-to-integrated flux ratio relative to those of other sources in the mosaic as a function of S/N.
From this analysis we conclude that CID-42 is unresolved at a resolution of 0.9$\arcsec$ and that it is marginally resolved at 0.7$\arcsec$. We report the peak flux density of CID-42, drawn from the SPW stack map to be $33.2 \pm 4.8 \,\mu$Jy and from the MSMF map $32.5 \pm 4.6 \,\mu$Jy. Since CID-42 is only  marginally resolved in the MSMF map, we report its integrated flux density of $39.7 \pm 9.5\,\mu$Jy, but we do not use this value in the spectral analysis. The error of peak flux density is determined from the local rms
 obtained via fitting the histogram of the pixel flux density values with a Gaussian, and the error for the integrated flux density is drawn from the elliptical Gaussian fit.

As the resolution of our 3~GHz data is around a factor of 2 higher than that of the large VLA-COSMOS 1.4~GHz data (\citealt{schinnerer07}), to test whether we may be out-resolving a fraction of the flux density at 3~GHz we used two methods to match the resolution of the 3~GHz map to the 1.4~GHz map. Firstly, we convolved the 3~GHz map to a resolution of 1.5$\arcsec$, matching that of the 1.4~GHz map.
Peak flux density of CID-42 in this map is  $36.8 \pm 7.3\,\mu$Jy\,beam$^{-1}$.
Secondly, in the cleaning process, we weighted the visibilities with the Gaussian taper in the uv-plane.
This approach gave a peak flux density of $ 34.6 \pm 6.3 \,\mu$Jy\,beam$^{-1}$.   
Both methods result in flux densities consistent with the higher resolution map within the error bars.

\section{Radio data analysis}
\label{sec:radiocid}
\subsection{Image analysis}
\begin{figure}
\includegraphics[ width=\columnwidth]{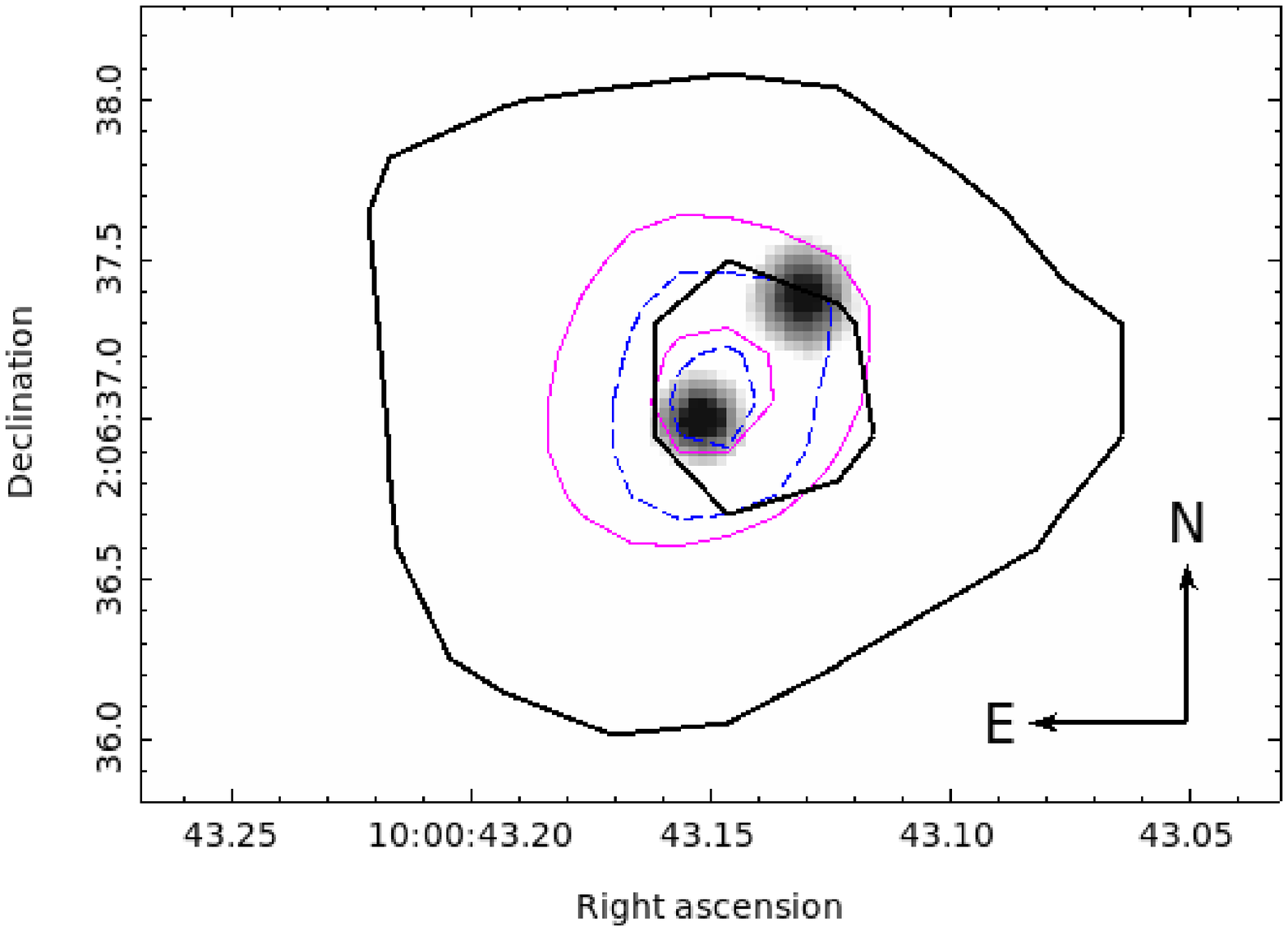}

\includegraphics[ width=\columnwidth]{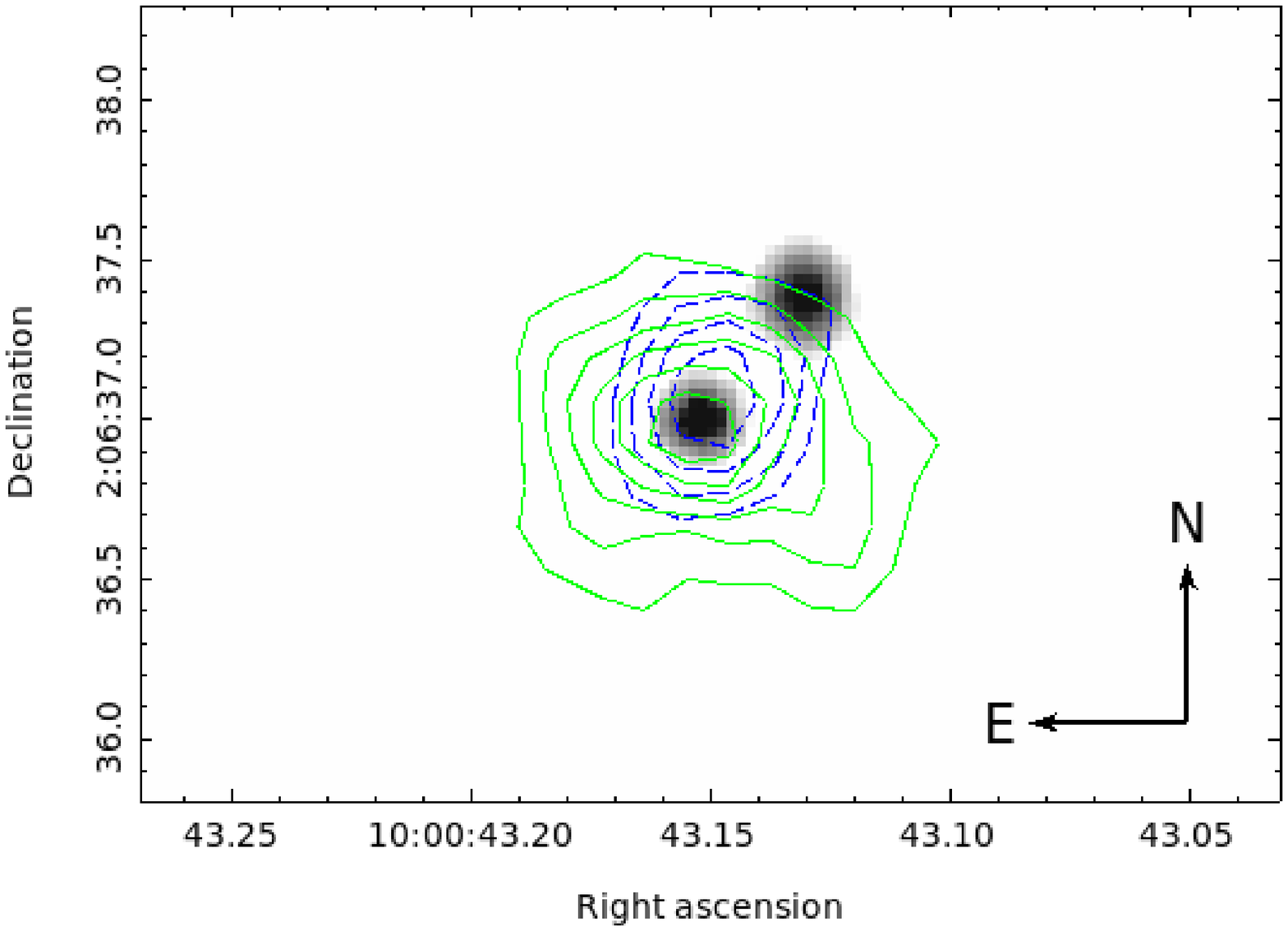}

\caption{ Top: HST/ACS gray-scale image showing the two optical components of CID-42 with contours from the radio data overlaid. For each radio map, two contour lines are shown that correspond to 3$\sigma$ and 6$\sigma$.
Thick black lines show data from the VLA 1.4~GHz large map from \protect\cite{schinnerer07} with $\sigma=10\,\mu$Jy\,beam$^{-1}$. Thin magenta lines and dashed blue lines are from the VLA 3~GHz SPW stack map and MSMF map with the local rms of $4.8\,\mu$Jy\,beam$^{-1}$ and $4.6\,\mu$Jy\,beam$^{-1}$, respectively. Bottom: HST/ACS gray-scale image of CID-42 with contours from the adaptively smoothed X-ray \chandra\ High Resolution Camera image with a 3 pixel radius Gaussian kernel by \protect\cite{civano12} in green and VLA 3~GHz MSMF map with 1$\sigma$ steps starting from the 3$\sigma$ in blue.}
 \label{fig:hstcont}
\end{figure}

Table \ref{tab:radio} lists all the radio data used for the CID-42 analysis presented here, their resolutions and rms in the corresponding maps.  In the top panel of Figure \ref{fig:hstcont} we show the HST/ACS image \citep{koekemoer07} of CID-42 with contours from the 1.4~GHz (black) and 3~GHz (magenta and blue) radio data overlaid. The resolution of 1.5$\arcsec$ at 1.4~GHz (\citealt{schinnerer07}) is not accurate enough to distinguish whether the radio emission is associated with the NW or the SE optical source (see also \citealt{civano10}). However, the 3~GHz data presented here at a resolution of 0.7$\arcsec$ show that the 6$\sigma$ contours are coincident with the SE component. The peak of the 3~GHz radio emission is located at
$\alpha_{2000.0}=10^{\rmn{h}} 00^{\rmn{m}} 43\fs 148, \delta_{2000.0}=+2\degr 06\arcmin 37\farcs 06$. 
The offset between the center of the SE component in the HST image and the center of the VLA 3~GHz emission is 0.08$\arcsec$.
This value is within the positional error of 0.1$\arcsec$ estimated from the ratio of resolution and S/N which dominates the positional uncertainty of faint sources. 
We compared the positions of 30 sources measured with the Very Long Baseline Array (VLBA) by Herrera Ruiz et al. (in prep.) with our 3~GHz mosaic and found that positional errors in our mosaic are less than 0.03$\arcsec$. This constrains the positional error of bright sources (S/N between 40 and 900) inside our mosaic. The low S/N sources positional error is therefore dominated by the mentioned term.

In the bottom panel of Figure \ref{fig:hstcont} we show that radio emission at 3~GHz is coincident with the X-ray emission (\citealt{civano12}). As described in detail in Section \ref{sec:sizes}, CID-42 is unresolved at a resolution of 0.9$\arcsec$ meaning that the FWHM 3~GHz radio emission is constrained to a region $<$ 4.5 kpc (at the source redshift) centered at the SE optical component. Since CID-42 is marginally resolved in the MSMF map at a resolution of 0.7$\arcsec$ we calculated its deconvolved size. The deconvolved FWHM major axis is 0.6$\arcsec$ and minor axis 0.1$\arcsec$ (corresponding to 3 kpc and 0.5 kpc respectively) with a position angle of -36$^{\circ}$ measured from North through East which could indicate an elongated shape of the source. However, better resolution and deeper data are needed to confirm this.

\begin{table*}

	\caption{Radio data used for the CID-42 flux density analysis along with the corresponding resolution and local rms noise inside the maps. 
	CID-42 is resolved in the 1.4~GHz VLA map and marginally resolved in the 3~GHz MSMF map where integrated flux density is also reported in parentheses, for all other maps peak flux density is reported.}

	\begin{center}
	\begin{tabular}{|c|c|c|c|c|c|}
	\hline
	$\nu$ (GHz) & Instrument & Resolution & RMS ($\mu$Jy\,beam$^{-1}$) & Flux density ($\mu$Jy\,beam$^{-1}$) & Author\\
	\hline
	0.32 & VLA   & 6$\arcsec$          & 442 & $< 1326$ & \protect\cite{smolcic2014} \\
	0.325 & GMRT  & 10.8$\arcsec$      & 80  & $ 240 \pm 80 $ & Karim et al. (in prep.) \\
	1.4  & VLA   & 1.5$\arcsec$        & 10 & $ 85 \pm 10$ &  \protect\cite{schinnerer07} \\
	  &    &         &  & ($138 \pm 38 \,\mu$Jy) &   \\
	1.5  & VLBA  & 0.015$\arcsec$      & 13 & $< 78 $ & \protect\cite{wrobel14} \\
	2.7  & VLA (SPW stack - low) & 0.9$\arcsec$   &  6.9 & $31.2 \pm 6.9$ & this paper\\
	3.4  & VLA (SPW stack - high) & 0.9$\arcsec$   & 6.5 & $35.4 \pm 6.5$ & this paper \\
	3.1  & VLA (SPW stack - full) & 0.9$\arcsec$   & 4.8 & $33.2 \pm 4.8$ & this paper\\
	3    & VLA (MSMF) & 0.7$\arcsec$  & 4.6 & $32.5 \pm 4.6$ & this paper\\
	   &  &   &  & ($39.7 \pm 9.5\,\mu$Jy) &  \\
	9    & VLA  & 0.32$\arcsec$       & 6 & $< 18 $ & \protect\cite{wrobel14}\\
	\hline
	\end{tabular} 
	\end{center}

	 \label{tab:radio}
\end{table*}

\subsection{Spectral analysis}
\begin{figure}
\includegraphics[ width=\columnwidth]{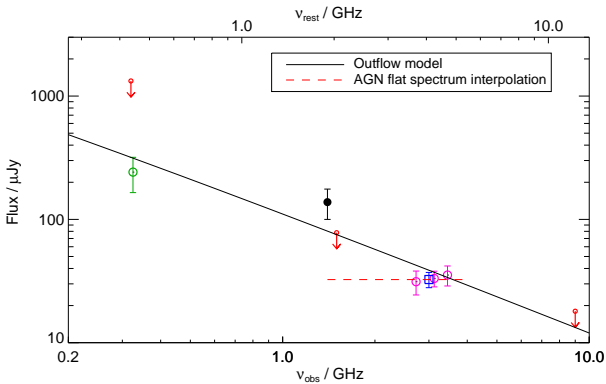}
\caption{ Radio spectrum of CID-42. Values are listed in Table \ref{tab:radio}.
Black point marks the total integrated flux density from the 1.4~GHz VLA data \protect\citep{schinnerer07}. 
Magenta points are peak flux densities from SPWs stacks from our new VLA 3~GHz data. Flux density from the MSMF cleaned map is shown with a blue square.
Green point is peak flux density from the GMRT map (Karim et al., in prep).
Red arrows show upper $3\sigma$ limit from the 320~MHz VLA (\protect\citealt{smolcic2014}), $3\sigma$ limit from the  9~GHz VLA data, and $6\sigma$ limit from the 1.5GHz VLBA data (\protect\citealt{wrobel14}). The flat spectrum interpolation for AGN is plotted with red dashed line. Black line corresponds to a shock generated by outflow model (see text for details).}
 \label{fig:cidfluxes}
\end{figure}

Figure \ref{fig:cidfluxes} shows spectral features of CID-42 across a frequency range of 320~MHz to 9~GHz. 
\cite{schinnerer07} observed CID-42 at 1.4~GHz and 1.5$\arcsec$ resolution within the VLA-COSMOS survey. They find the source resolved at that resolution and report an integrated flux density of 138 $\pm$ 38 $\mu$Jy.
\cite{wrobel14} found no detection using a high resolution of 0.015$\arcsec$ VLBA at 1.5~GHz and report a $6\sigma$ upper limit of $78\,\mu$Jy\,beam$^{-1}$. \cite{wrobel14} also did not detect the source at the higher frequency of 9~GHz using the VLA with a $3\sigma$ upper limit of $18\,\mu$Jy\,beam$^{-1}$.
In Figure \ref{fig:cidfluxes}  we further plot the 325 MHz flux density of 240 $\pm$ 80 $\mu$Jy\,beam$^{-1}$ from the Giant Metrewave Radio Telescope (GMRT) data (Karim et al. in prep.). The GMRT gives a  tentative $3\sigma$ detection for CID-42, but since we know the source location and we expect that it is unresolved at a resolution of $\approx 10\arcsec$ this may justify a peak flux density measurement. For this reason the spectrum shown in Figure \ref{fig:cidfluxes}  is not strongly constrained by the 325~MHz measurement.

In the 2~GHz bandwidth centered at 3~GHz, the galaxy shows a slightly inverted spectra. This behaviour is evident in the flux densities extracted from the SPW stacks (see Figure \ref{fig:cidfluxes}) and also from the spectral index calculated in the MSMF alpha map where $\alpha=0.4\pm 0.3$. For the sake of simplicity, we will treat this as a flat spectrum ($\alpha=0$). 
Using only the MSMF value, the flux density of CID-42 at 3~GHz corresponds to rest frame spectral luminosity of $L_{\text{3~GHz}}\approx 1 \times 10^{22} $\, W\,Hz$^{-1}$. 
Based on radio-loud vs. radio-quiet definitions in literature relying only on radio luminosity values (e.g. \citealt{miller1990}, see also \citealt{Balokovic2012}), this falls into the radio-quiet regime.

\begin{figure*}
\includegraphics[ width=0.9\columnwidth]{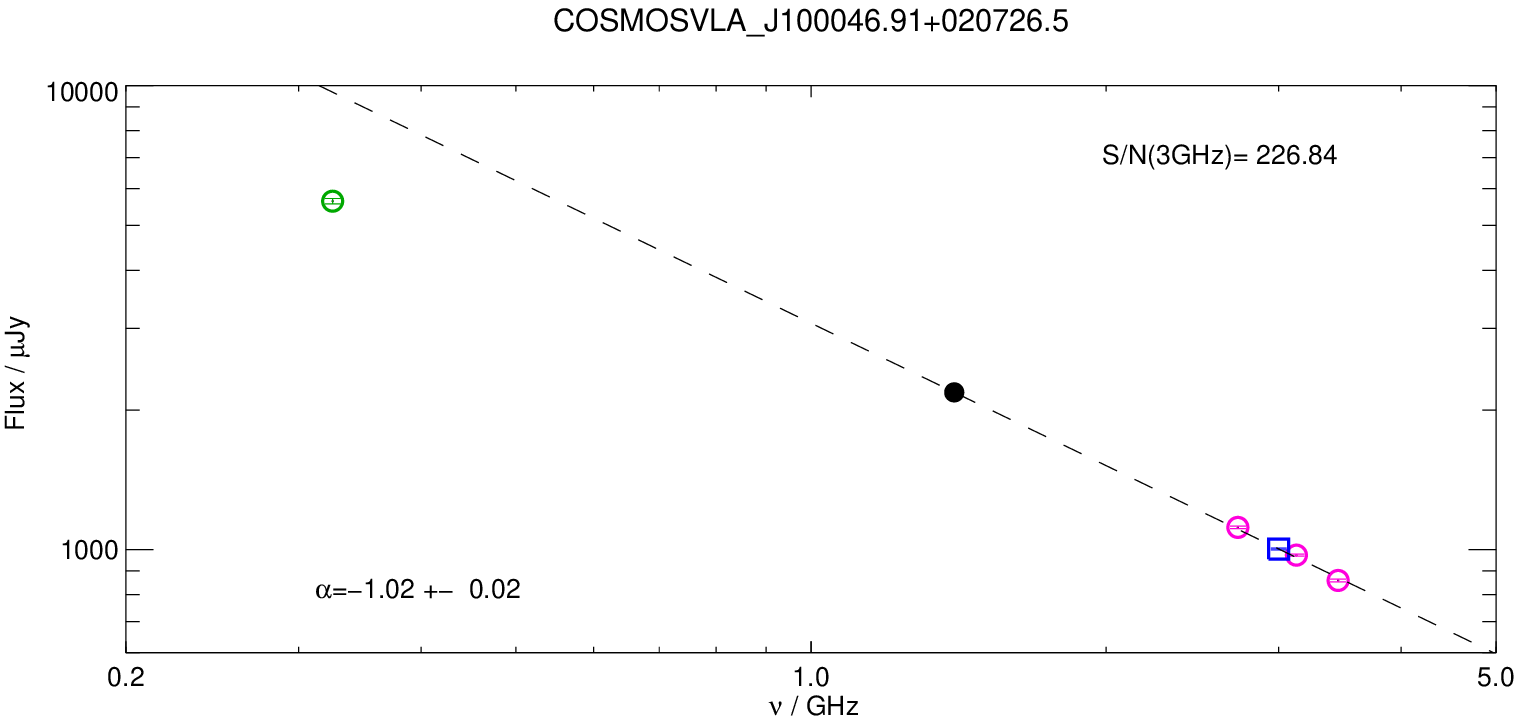}
\includegraphics[ width=0.9\columnwidth]{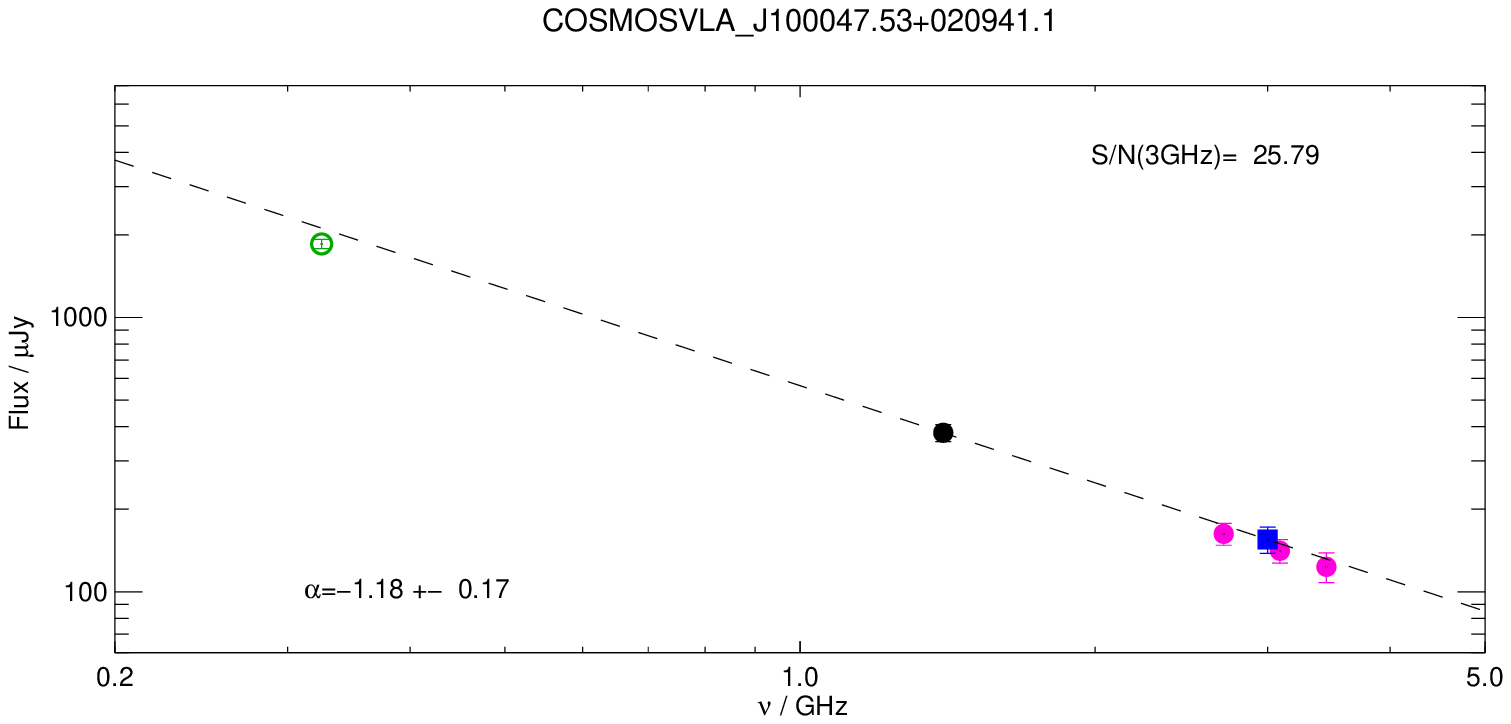}

\includegraphics[ width=0.9\columnwidth]{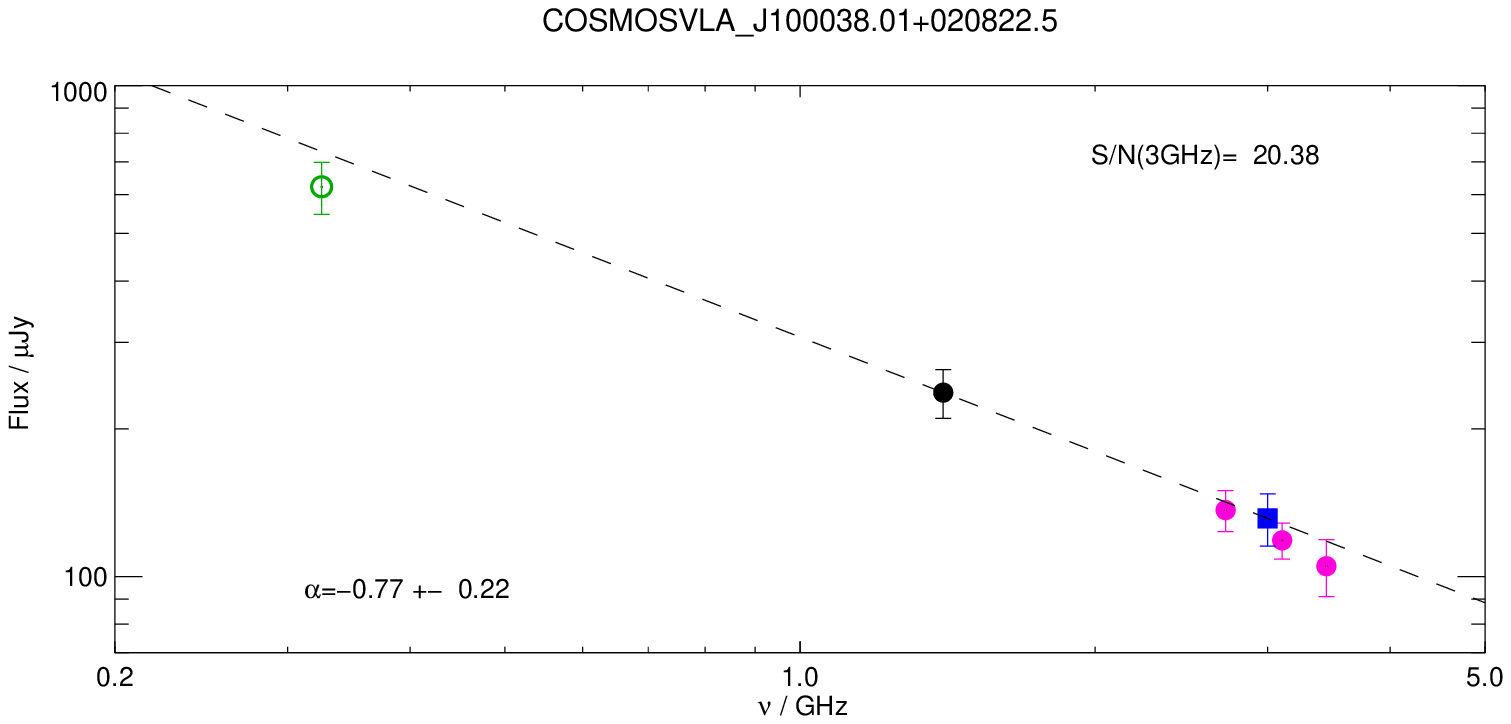}
\includegraphics[ width=0.9\columnwidth]{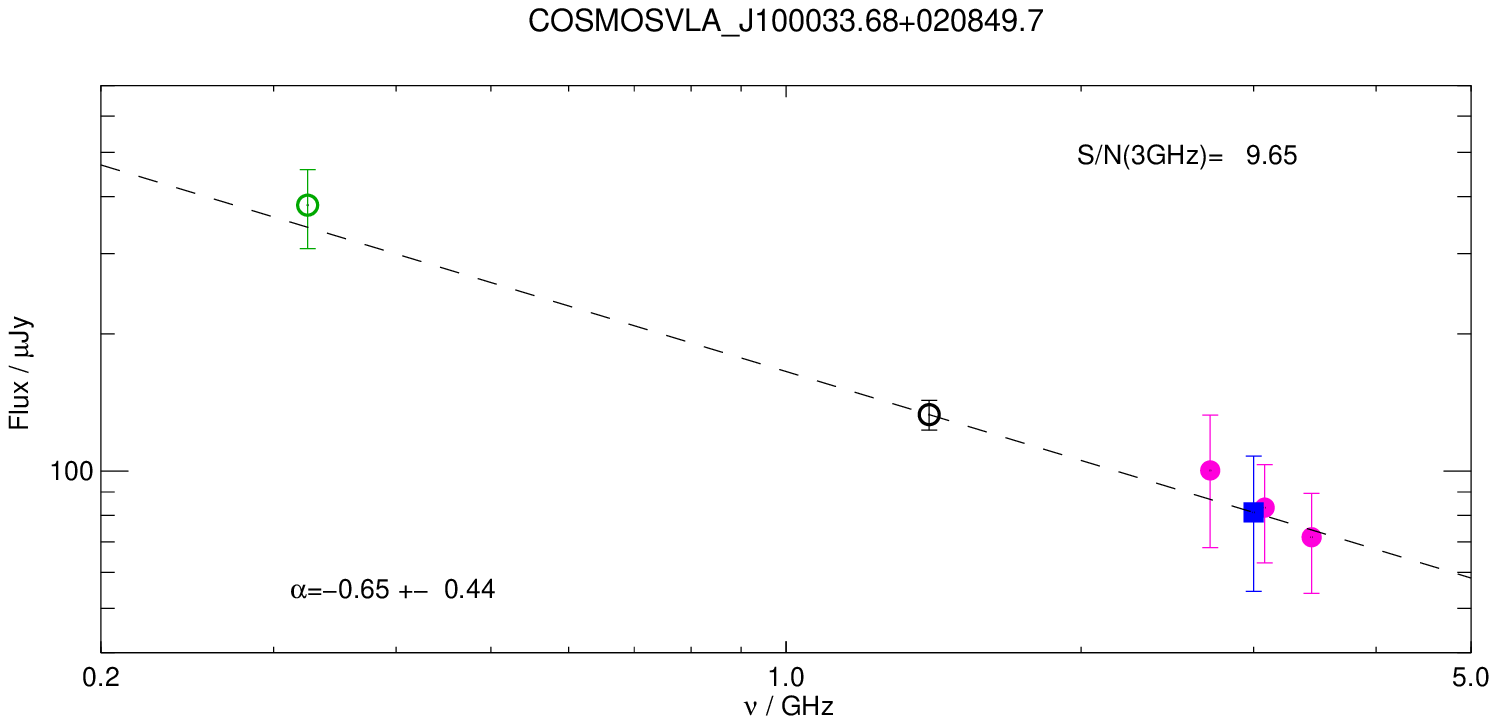}

\includegraphics[ width=0.9\columnwidth]{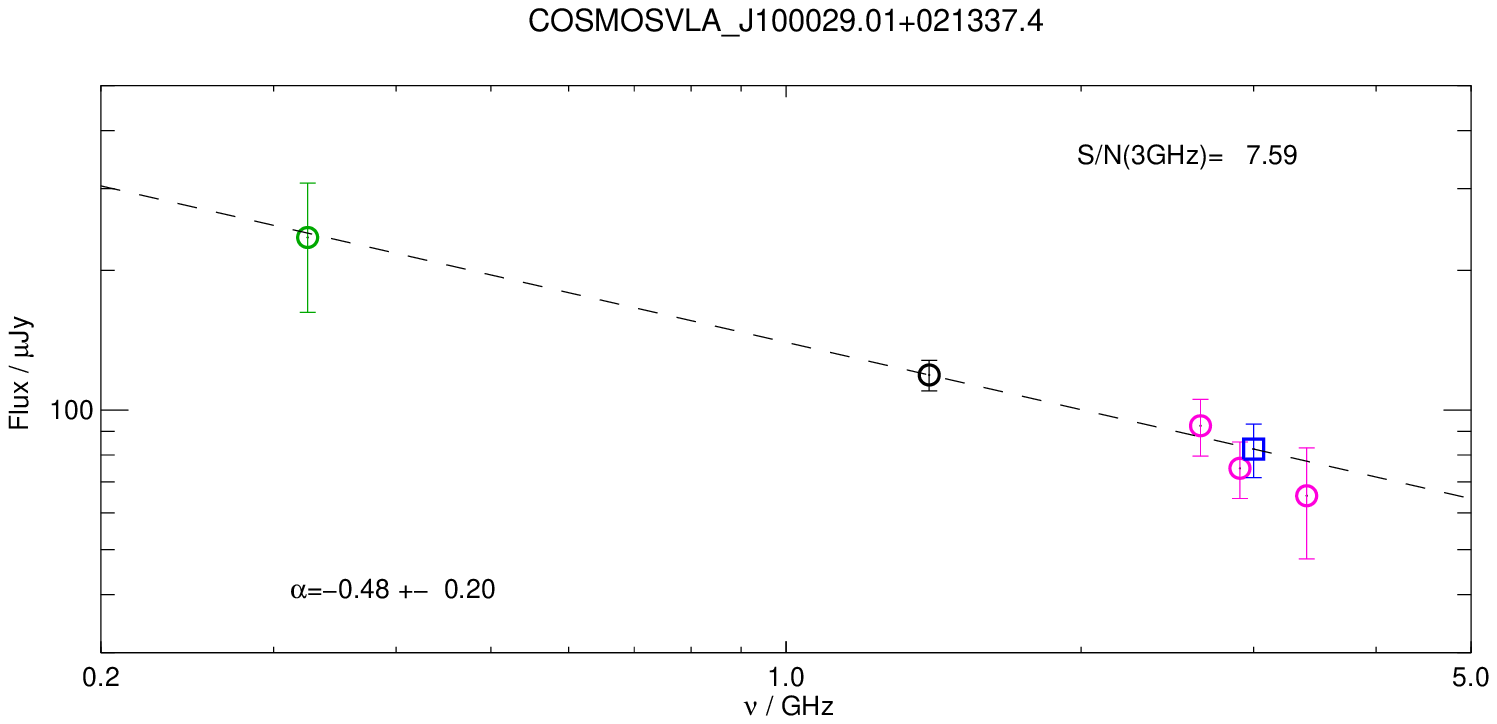}
\includegraphics[ width=0.9\columnwidth]{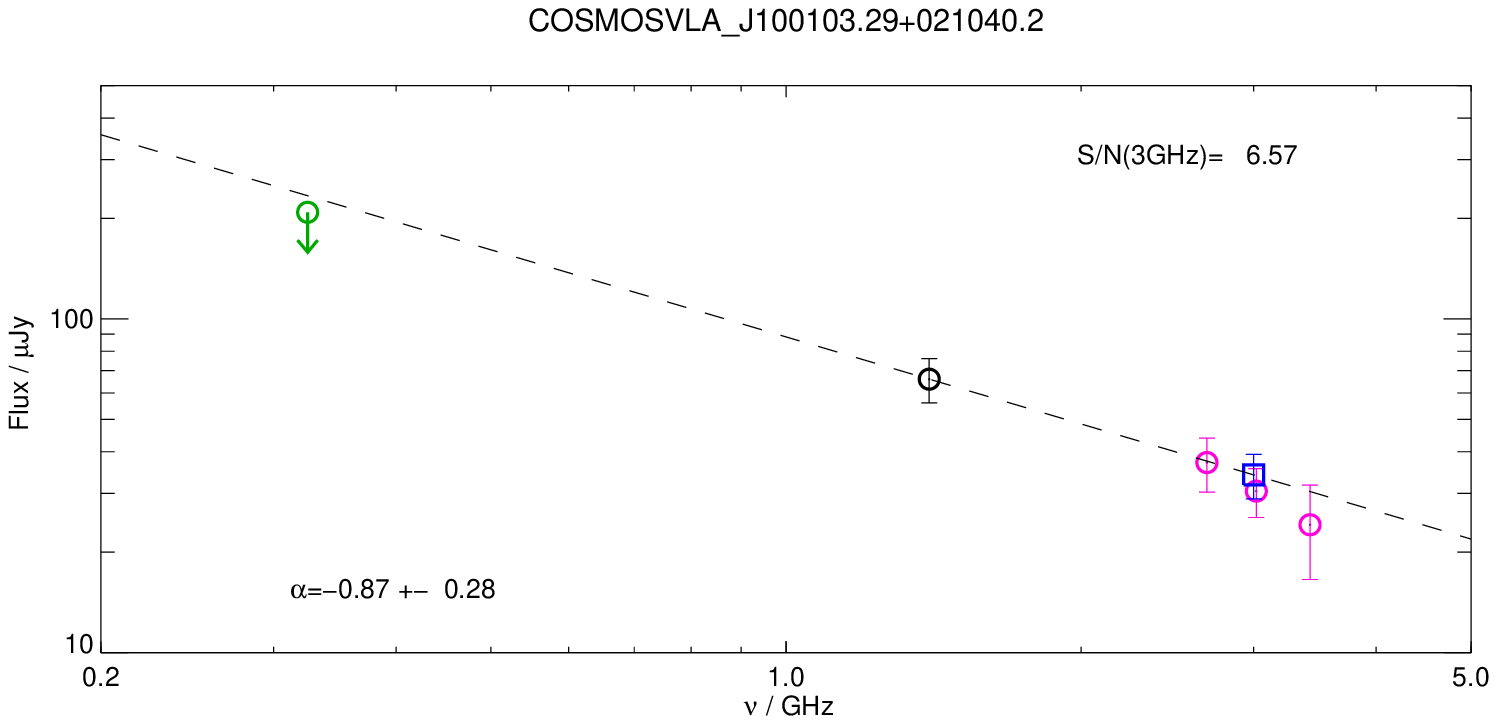}

\includegraphics[ width=0.9\columnwidth]{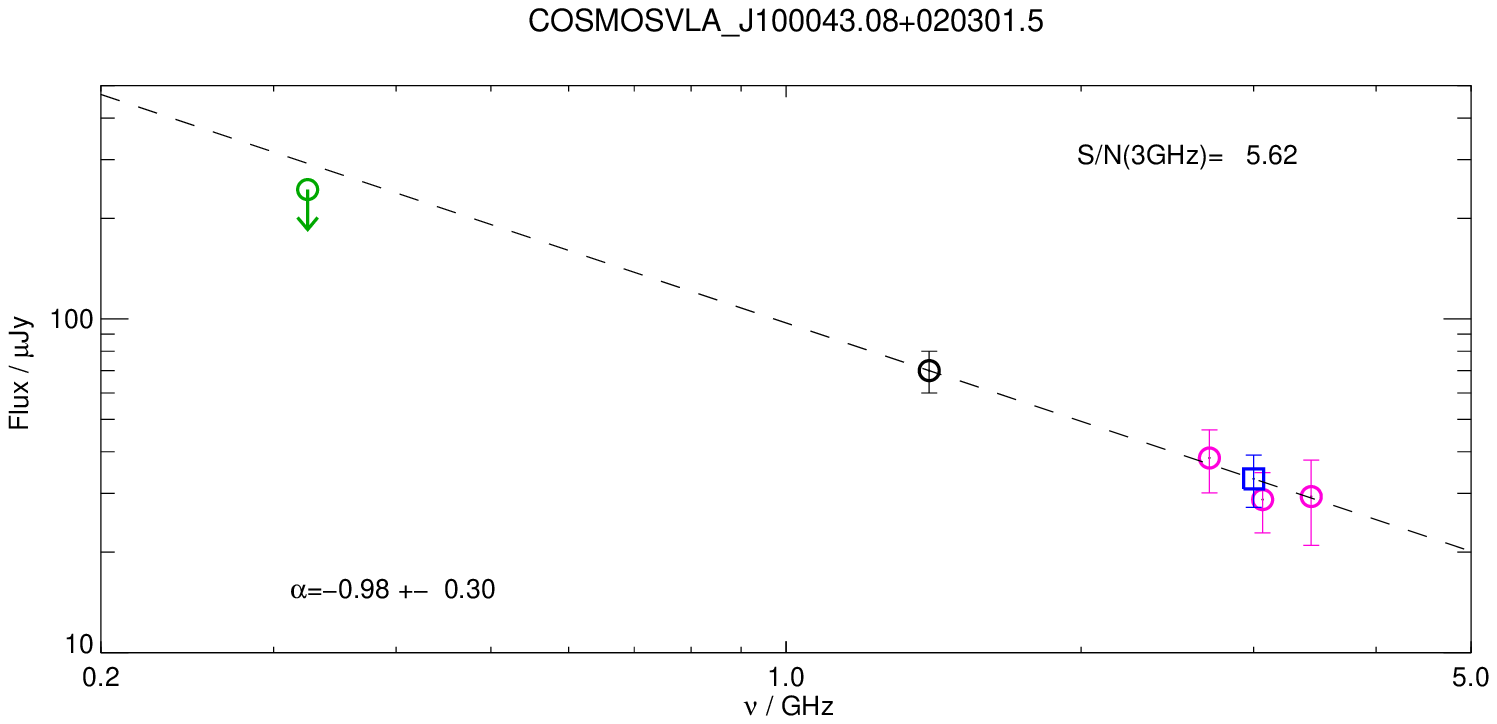}
\includegraphics[ width=0.9\columnwidth]{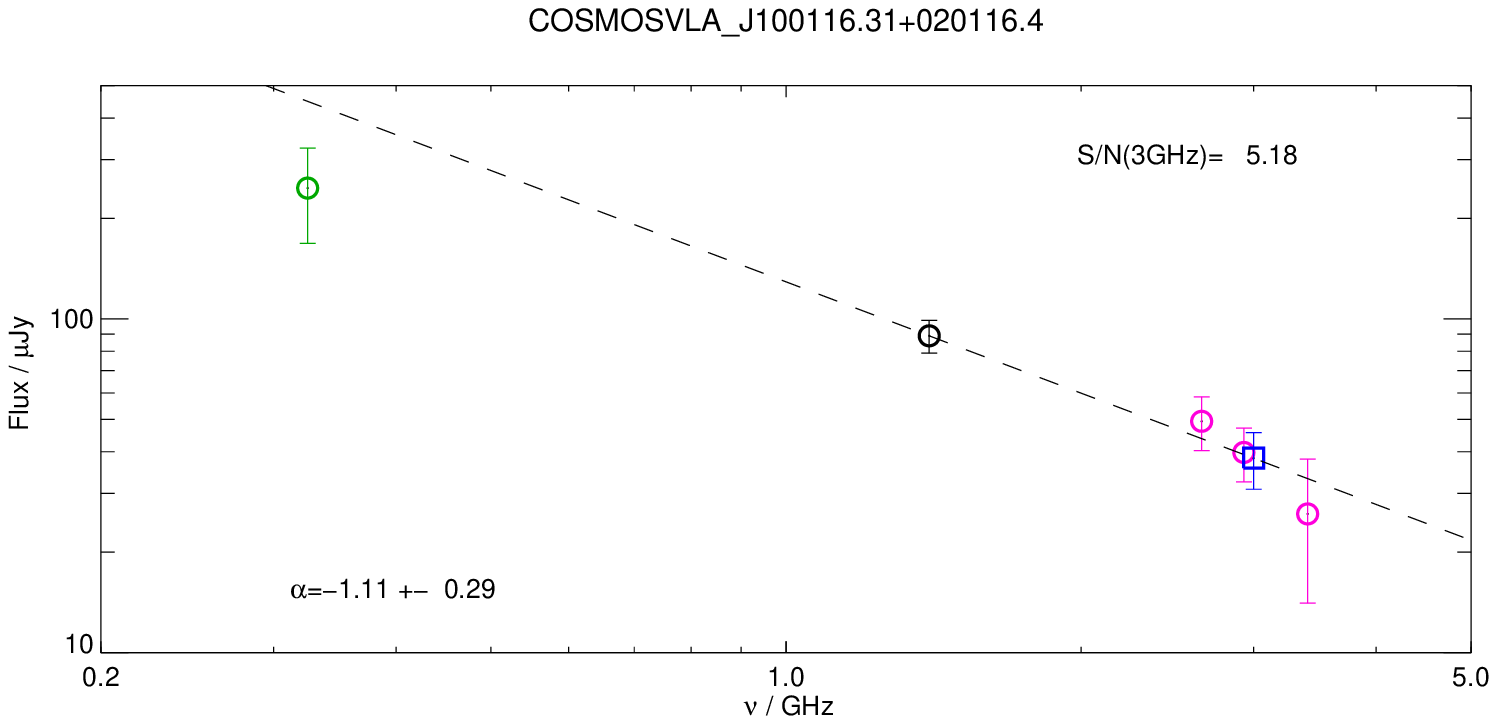}

\caption{ Flux densities for selected sources as a function of frequency inside the VLA-COSMOS pointing containing CID-42 at different S/N ratios. Filled symbols are used for integrated flux densities of resolved sources and empty symbols for peak flux densities of unresolved sources. 
Black points mark the flux density from the 1.4~GHz large project catalogue \protect\citep{schinnerer07}. 
Magenta points are flux densities from the low frequency, whole bandwidth, and high frequency SPW stack from the VLA 3~GHz data.
Green points are peak flux densities from the GMRT map (Karim et al., in prep).
 Flux density from the MSMF cleaned map is shown with a blue square.
  Black dashed line correspond to spectral index derived from the 1.4~GHz and 3~GHz point. 
   Error bars represent the local rms for peak flux densities and fit error for integrated flux densities.}
 \label{fig:s4}

\end{figure*}

To test the consistency between the two different wideband imaging methods and the robustness of our 3~GHz data in general and thus that for CID-42, we have closely examined the spectral behavior of 109 sources in the mosaic in different locations within the primary beam.
In Figure \ref{fig:s4}, we show eight representative sources over a broad S/N range (from 230 to 5). 
 In the spectral plots we also show our 3~GHz flux densities with the 1.4~GHz Large Project catalog values \citep{schinnerer07} and the 325~MHz GMRT data (Karim et al. in prep.).
We find that both imaging methods (SPW stack and MSMF) give consistent flux densities, also consistent with the flux densities at lower frequencies.
Thus, we conclude that our map is robust with no systematics for sources at either high or low S/N. 

\section{Discussion}

The unresolved emission at 0.9$\arcsec$ and flat spectrum we derived here for CID-42 at 3~GHz (see Figure \ref{fig:cidfluxes}) suggests that the  entire observed radio emission at 3~GHz (with 2~GHz bandwidth) may arise from an AGN within the SE component.
The VLBA upper limit at 1.5~GHz by \cite{wrobel14}  is consistent with the flat spectrum AGN picture, as the VLBA can observe scales up to 150 mas (750 pc) but it is not sensitive to star formation induced emission.
The non detection at 9~GHz \citep{wrobel14} could be explained by steepening of the spectrum due to synchrotron losses at higher frequencies.
The 1.4~GHz detection in the VLA-COSMOS survey \citep{schinnerer07} is resolved at a coarse resolution of 1.5$\arcsec$ and exhibits significantly higher flux density than the 3~GHz and the VLBA flux densities. This may  suggest that the 1.4~GHz flux density arises from a region more extended than 0.9$\arcsec$ (4.5 kpc) observed at 3~GHz.

Assuming, as argued above, that the 3~GHz emission is dominated by a flat spectrum AGN, by extrapolating this flux density to 1.4~GHz and comparing it with the observed one at that frequency we find an excess of $\approx$~105~$\mu$Jy (see Figure \ref{fig:cidfluxes}). 
As tested in Section \ref{sec:sizes} we found no evidence that this may be a resolution effect. Thus, if this excess were due to star-formation than it would need to have  a spectral slope of $\alpha < -2$ to account for the observed 3~GHz emission. Such a steep slope is highly untypical for star-formation as shown by e.g. \cite{kimball08} who find that typical spectral slopes for star-forming galaxies are between -0.5 and -1.5 based on SDSS-NVSS-FIRST data. The steep spectrum may possibly reflect aging of the electrons by synchrotron radiation 
%and inverse Compton scattering on the cosmic microwave background (CMB) 
with a characteristic break frequency between 1.4 and 3~GHz (see \citealt{laan1969} and \citealt{miley1980}) or could occur in radio jets. 
We have further tried to model the full observed spectrum shown in Figure \ref{fig:cidfluxes} assuming bow shock and mass outflow models generated by an outflow from the SMBH (for details see \citealt{loeb07} and  \citealt{wang2014}). Both models are able to fit the data with a resulting spectral slope of $\alpha \approx -1$ in the range from 320~MHz to 9~GHz, yet the fits still cannot account for the excess at 1.4~GHz rendering the physical interpretation difficult (the outflow model is plotted in Figure \ref{fig:cidfluxes}).
It is possible that CID-42 is a variable source in radio. We can get an insight into the source variability by comparing the various radio COSMOS data. We find that CID-42 is not showing flux density variations in the 1.4~GHz VLA-COSMOS data taken between the years 2003 and 2006 (\citealt{schinnerer04,schinnerer07,schinnerer10}). If variability is the reason for the 1.4~GHz radio excess of CID-42, it would have had to occur between years 2006 and 2013 or on shorter timescales (day/month).
To test whether CID-42 is variable on monthly scales we concatenated our 3~GHz A-array observations into two epochs lasting approximately 20 days each and imaged them separately using the MSMF algorithm. The peak flux density extracted from the first epoch (28 November 2012 to 19 December 2012) is $34.6 \pm 6.7\,\mu$Jy and from the second epoch (20 December 2012 to 7 January 2013) is $24.8 \pm 6.3 \,\mu$Jy (errors represent local rms in the maps). 
Although the flux densities agree withing $2\sigma$, their difference between these two time periods could indicate that CID-42 is variable on monthly time scales. Further observations are needed to confirm this.

In summary, the SE component of CID-42 appears to be a flat spectrum radio-quiet AGN with associated extended emission, the source of which is still unclear. On the other hand, the NW component is not detected in our 3~GHz data.
If the NW component is an obscured AGN, as already suggested by in \cite{civano10} and \cite{wrobel14}, then it is also radio-quiet with a $3\sigma$ upper limit rest frame spectral luminosity at 3~GHz of $L_{\text{3~GHz}}<5.6 \times 10^{21} $\, W\,Hz$^{-1}$ (assuming a spectral index of -0.8). 
The results presented here are not inconsistent with the recoiling black hole scenario, but still cannot rule out the presence of an obscured radio quiet SMBH in the NW component.
Further spectroscopic observations would be able to spatially resolve the two nuclei in the optical spectrum.

\section{Summary}
\label{sec:summary}

We used the first 130 hours of VLA-COSMOS 3~GHz data to analyze the radio synchrotron spectrum of CID-42, the best candidate recoiling black hole in the COSMOS field. Due to the large 2~GHz bandwidth, we imaged the data with two different methods: spectral window (SPW) stacking and multi-scale multi-frequency (MSMF). Both of these methods gave consistent flux densities for CID-42. Our $7\sigma$ detection shows that all of the 3~GHz radio emission is arising from the SE component of CID-42 and our new 3~GHz radio data confirm that the SE component is  an unobscured type I radio-quiet AGN. These data combined with other radio data from the literature (\citealt{schinnerer07}, \citealt{wrobel14}) suggest that the radio emission is composed of a flat spectrum AGN core and perhaps a more extended region of an aged electron synchrotron population or shocks generated by an outflow from the black hole. Only an upper limit of radio emission can be given for the NW component. 
There are indications that CID-42 could be variable on monthly scales based on the $2\sigma$ flux density variation between two different time epochs but further observations are needed to confirm this.

\section*{acknowledgement}
We thank the anonymous referee for providing helpful comments which improved the paper.
This research was funded by the European Union’s Seventh Frame-work program under grant agreement 337595 (ERC Starting Grant, ’CoSMass’).
The National Radio Astronomy Observatory is a facility of the National Science Foundation operated under cooperative agreement by Associated Universities, Inc.

\bibliographystyle{mn2e}
\bibliography{reflist}

\begin{thebibliography}{48}
\expandafter\ifx\csname natexlab\endcsname\relax\def\natexlab#1{#1}\fi

\bibitem[{{Balokovi{\'c}} {et~al}\mbox{.}(2012){Balokovi{\'c}}, {Smol{\v
  c}i{\'c}}, {Ivezi{\'c}}, {Zamorani}, {Schinnerer}, \&
  {Kelly}}]{Balokovic2012}
{Balokovi{\'c}} M., {Smol{\v c}i{\'c}} V., {Ivezi{\'c}} {\v Z}., {Zamorani} G.,
  {Schinnerer} E., {Kelly} B.~C., 2012, \apj, 759, 30

\bibitem[{{Batcheldor} {et~al}\mbox{.}(2010){Batcheldor}, {Robinson}, {Axon},
  {Perlman}, \& {Merritt}}]{batcheldor10}
{Batcheldor} D., {Robinson} A., {Axon} D.~J., {Perlman} E.~S., {Merritt} D.,
  2010, \apjl, 717, L6

\bibitem[{{Bekenstein}(1973)}]{bekenstein73}
{Bekenstein} J.~D., 1973, \apj, 183, 657

\bibitem[{{Bianchi} {et~al}\mbox{.}(2013){Bianchi}, {Piconcelli},
  {P{\'e}rez-Torres}, {Fiore}, {La Franca}, {Mathur}, \& {Matt}}]{bianchi13}
{Bianchi} S., {Piconcelli} E., {P{\'e}rez-Torres} M.~{\'A}., {Fiore} F., {La
  Franca} F., {Mathur} S., {Matt} G., 2013, \mnras, 435, 2335

\bibitem[{{Blecha} {et~al}\mbox{.}(2013){Blecha}, {Civano}, {Elvis}, \&
  {Loeb}}]{blecha13}
{Blecha} L., {Civano} F., {Elvis} M., {Loeb} A., 2013, \mnras, 428, 1341

\bibitem[{{Blecha} {et~al}\mbox{.}(2011){Blecha}, {Cox}, {Loeb}, \&
  {Hernquist}}]{blecha11}
{Blecha} L., {Cox} T.~J., {Loeb} A., {Hernquist} L., 2011, \mnras, 412, 2154

\bibitem[{{Bondi} {et~al}\mbox{.}(2003){Bondi}, {Ciliegi}, {Zamorani},
  {Gregorini}, {Vettolani}, {Parma}, {de Ruiter}, {Le Fevre}, {Arnaboldi},
  {Guzzo}, {Maccagni}, {Scaramella}, {Adami}, {Bardelli}, {Bolzonella},
  {Bottini}, {Cappi}, {Foucaud}, {Franzetti}, {Garilli}, {Gwyn}, {Ilbert},
  {Iovino}, {Le Brun}, {Marano}, {Marinoni}, {McCracken}, {Meneux}, {Pollo},
  {Pozzetti}, {Radovich}, {Ripepi}, {Rizzo}, {Scodeggio}, {Tresse},
  {Zanichelli}, \& {Zucca}}]{bondi03}
{Bondi} M. {et~al.}, 2003, \aap, 403, 857

\bibitem[{{Bonning}, {Shields} \& {Salviander}(2007){Bonning}, {Shields}, \&
  {Salviander}}]{bonning07}
{Bonning} E.~W., {Shields} G.~A., {Salviander} S., 2007, \apjl, 666, L13

\bibitem[{{Bourke}, {Mooley} \& {Hallinan}(2014){Bourke}, {Mooley}, \&
  {Hallinan}}]{bourke2014proc}
{Bourke} S., {Mooley} K., {Hallinan} G., 2014, in Astronomical Society of the
  Pacific Conference Series, Vol. 485, Astronomical Society of the Pacific
  Conference Series, {Manset} N., {Forshay} P., eds., p. 367

\bibitem[{{Civano} {et~al}\mbox{.}(2012{\natexlab{a}}){Civano}, {Elvis},
  {Brusa}, {Comastri}, {Salvato}, {Zamorani}, {Aldcroft}, {Bongiorno}, {Capak},
  {Cappelluti}, {Cisternas}, {Fiore}, {Fruscione}, {Hao}, {Kartaltepe},
  {Koekemoer}, {Gilli}, {Impey}, {Lanzuisi}, {Lusso}, {Mainieri}, {Miyaji},
  {Lilly}, {Masters}, {Puccetti}, {Schawinski}, {Scoville}, {Silverman},
  {Trump}, {Urry}, {Vignali}, \& {Wright}}]{civano12b}
{Civano} F. {et~al.}, 2012{\natexlab{a}}, \apjs, 201, 30

\bibitem[{{Civano} {et~al}\mbox{.}(2012{\natexlab{b}}){Civano}, {Elvis},
  {Lanzuisi}, {Aldcroft}, {Trichas}, {Bongiorno}, {Brusa}, {Blecha},
  {Comastri}, {Loeb}, {Salvato}, {Fruscione}, {Koekemoer}, {Komossa}, {Gilli},
  {Mainieri}, {Piconcelli}, \& {Vignali}}]{civano12}
{Civano} F. {et~al.}, 2012{\natexlab{b}}, \apj, 752, 49

\bibitem[{{Civano} {et~al}\mbox{.}(2010){Civano}, {Elvis}, {Lanzuisi},
  {Jahnke}, {Zamorani}, {Blecha}, {Bongiorno}, {Brusa}, {Comastri}, {Hao},
  {Leauthaud}, {Loeb}, {Mainieri}, {Piconcelli}, {Salvato}, {Scoville},
  {Trump}, {Vignali}, {Aldcroft}, {Bolzonella}, {Bressert}, {Finoguenov},
  {Fruscione}, {Koekemoer}, {Cappelluti}, {Fiore}, {Giodini}, {Gilli}, {Impey},
  {Lilly}, {Lusso}, {Puccetti}, {Silverman}, {Aussel}, {Capak}, {Frayer}, {Le
  Floch}, {McCracken}, {Sanders}, {Schiminovich}, \& {Taniguchi}}]{civano10}
{Civano} F. {et~al.}, 2010, \apj, 717, 209

\bibitem[{{Colpi} \& {Dotti}(2009)}]{colpi09}
{Colpi} M., {Dotti} M., 2009, ArXiv e-prints

\bibitem[{{Comerford} {et~al}\mbox{.}(2009){Comerford}, {Griffith}, {Gerke},
  {Cooper}, {Newman}, {Davis}, \& {Stern}}]{comerford09}
{Comerford} J.~M., {Griffith} R.~L., {Gerke} B.~F., {Cooper} M.~C., {Newman}
  J.~A., {Davis} M., {Stern} D., 2009, \apjl, 702, L82

\bibitem[{{Condon} {et~al}\mbox{.}(2012){Condon}, {Cotton}, {Fomalont},
  {Kellermann}, {Miller}, {Perley}, {Scott}, {Vernstrom}, \& {Wall}}]{condon12}
{Condon} J.~J. {et~al.}, 2012, \apj, 758, 23

\bibitem[{{Elvis} {et~al}\mbox{.}(2009){Elvis}, {Civano}, {Vignali},
  {Puccetti}, {Fiore}, {Cappelluti}, {Aldcroft}, {Fruscione}, {Zamorani},
  {Comastri}, {Brusa}, {Gilli}, {Miyaji}, {Damiani}, {Koekemoer}, {Finoguenov},
  {Brunner}, {Urry}, {Silverman}, {Mainieri}, {Hasinger}, {Griffiths},
  {Carollo}, {Hao}, {Guzzo}, {Blain}, {Calzetti}, {Carilli}, {Capak}, {Ettori},
  {Fabbiano}, {Impey}, {Lilly}, {Mobasher}, {Rich}, {Salvato}, {Sanders},
  {Schinnerer}, {Scoville}, {Shopbell}, {Taylor}, {Taniguchi}, \&
  {Volonteri}}]{elvis09}
{Elvis} M. {et~al.}, 2009, \apjs, 184, 158

\bibitem[{{Eracleous} {et~al}\mbox{.}(2012){Eracleous}, {Boroson}, {Halpern},
  \& {Liu}}]{eracleous12}
{Eracleous} M., {Boroson} T.~A., {Halpern} J.~P., {Liu} J., 2012, \apjs, 201,
  23

\bibitem[{{Guedes} {et~al}\mbox{.}(2011){Guedes}, {Madau}, {Mayer}, \&
  {Callegari}}]{guedes11}
{Guedes} J., {Madau} P., {Mayer} L., {Callegari} S., 2011, \apj, 729, 125

\bibitem[{{Hopkins} {et~al}\mbox{.}(2008){Hopkins}, {Hernquist}, {Cox}, \&
  {Kere{\v s}}}]{hopkins08}
{Hopkins} P.~F., {Hernquist} L., {Cox} T.~J., {Kere{\v s}} D., 2008, \apjs,
  175, 356

\bibitem[{{Jonker} {et~al}\mbox{.}(2010){Jonker}, {Torres}, {Fabian}, {Heida},
  {Miniutti}, \& {Pooley}}]{jonker10}
{Jonker} P.~G., {Torres} M.~A.~P., {Fabian} A.~C., {Heida} M., {Miniutti} G.,
  {Pooley} D., 2010, \mnras, 407, 645

\bibitem[{{Kimball} \& {Ivezi{\'c}}(2008)}]{kimball08}
{Kimball} A.~E., {Ivezi{\'c}} {\v Z}., 2008, \aj, 136, 684

\bibitem[{{Koekemoer} {et~al}\mbox{.}(2007){Koekemoer}, {Aussel}, {Calzetti},
  {Capak}, {Giavalisco}, {Kneib}, {Leauthaud}, {Le F{\`e}vre}, {McCracken},
  {Massey}, {Mobasher}, {Rhodes}, {Scoville}, \& {Shopbell}}]{koekemoer07}
{Koekemoer} A.~M. {et~al.}, 2007, \apjs, 172, 196

\bibitem[{{Komossa}(2012)}]{komossa12}
{Komossa} S., 2012, Advances in Astronomy, 2012

\bibitem[{{Komossa}, {Zhou} \& {Lu}(2008){Komossa}, {Zhou}, \&
  {Lu}}]{komossa08}
{Komossa} S., {Zhou} H., {Lu} H., 2008, \apjl, 678, L81

\bibitem[{{Koss} {et~al}\mbox{.}(2014){Koss}, {Blecha}, {Mushotzky}, {Hung},
  {Veilleux}, {Trakhtenbrot}, {Schawinski}, {Stern}, {Smith}, {Li}, {Man},
  {Filippenko}, {Mauerhan}, {Stanek}, \& {Sanders}}]{koss14}
{Koss} M. {et~al.}, 2014, ArXiv e-prints

\bibitem[{{Larson} {et~al}\mbox{.}(2011){Larson}, {Dunkley}, {Hinshaw},
  {Komatsu}, {Nolta}, {Bennett}, {Gold}, {Halpern}, {Hill}, {Jarosik}, {Kogut},
  {Limon}, {Meyer}, {Odegard}, {Page}, {Smith}, {Spergel}, {Tucker}, {Weiland},
  {Wollack}, \& {Wright}}]{larson2011}
{Larson} D. {et~al.}, 2011, \apjs, 192, 16

\bibitem[{{Loeb}(2007)}]{loeb07}
{Loeb} A., 2007, Physical Review Letters, 99, 041103

\bibitem[{{Merritt} {et~al}\mbox{.}(2006){Merritt}, {Storchi-Bergmann},
  {Robinson}, {Batcheldor}, {Axon}, \& {Cid Fernandes}}]{merritt06}
{Merritt} D., {Storchi-Bergmann} T., {Robinson} A., {Batcheldor} D., {Axon} D.,
  {Cid Fernandes} R., 2006, \mnras, 367, 1746

\bibitem[{{Miley}(1980)}]{miley1980}
{Miley} G., 1980, \araa, 18, 165

\bibitem[{{Miller}, {Peacock} \& {Mead}(1990){Miller}, {Peacock}, \&
  {Mead}}]{miller1990}
{Miller} L., {Peacock} J.~A., {Mead} A.~R.~G., 1990, \mnras, 244, 207

\bibitem[{Peres(1962)}]{peres62}
Peres A., 1962, Phys. Rev., 128, 2471

\bibitem[{{Rau}, {Bhatnagar} \& {Owen}(2014){Rau}, {Bhatnagar}, \&
  {Owen}}]{rau14}
{Rau} U., {Bhatnagar} S., {Owen} F.~N., 2014, ArXiv e-prints

\bibitem[{{Rau} \& {Cornwell}(2011)}]{rau11}
{Rau} U., {Cornwell} T.~J., 2011, \aap, 532, A71

\bibitem[{{Robinson} {et~al}\mbox{.}(2010){Robinson}, {Young}, {Axon}, {Kharb},
  \& {Smith}}]{robinson10}
{Robinson} A., {Young} S., {Axon} D.~J., {Kharb} P., {Smith} J.~E., 2010,
  \apjl, 717, L122

\bibitem[{{Schinnerer} {et~al}\mbox{.}(2004){Schinnerer}, {Carilli},
  {Scoville}, {Bondi}, {Ciliegi}, {Vettolani}, {Le F{\`e}vre}, {Koekemoer},
  {Bertoldi}, \& {Impey}}]{schinnerer04}
{Schinnerer} E. {et~al.}, 2004, \aj, 128, 1974

\bibitem[{{Schinnerer} {et~al}\mbox{.}(2010){Schinnerer}, {Sargent}, {Bondi},
  {Smol{\v c}i{\'c}}, {Datta}, {Carilli}, {Bertoldi}, {Blain}, {Ciliegi},
  {Koekemoer}, \& {Scoville}}]{schinnerer10}
{Schinnerer} E. {et~al.}, 2010, \apjs, 188, 384

\bibitem[{{Schinnerer} {et~al}\mbox{.}(2007){Schinnerer}, {Smol{\v c}i{\'c}},
  {Carilli}, {Bondi}, {Ciliegi}, {Jahnke}, {Scoville}, {Aussel}, {Bertoldi},
  {Blain}, {Impey}, {Koekemoer}, {Le Fevre}, \& {Urry}}]{schinnerer07}
{Schinnerer} E. {et~al.}, 2007, \apjs, 172, 46

\bibitem[{{Scoville} {et~al}\mbox{.}(2007){Scoville}, {Aussel}, {Brusa},
  {Capak}, {Carollo}, {Elvis}, {Giavalisco}, {Guzzo}, {Hasinger}, {Impey},
  {Kneib}, {LeFevre}, {Lilly}, {Mobasher}, {Renzini}, {Rich}, {Sanders},
  {Schinnerer}, {Schminovich}, {Shopbell}, {Taniguchi}, \&
  {Tyson}}]{scoville07a}
{Scoville} N. {et~al.}, 2007, \apjs, 172, 1

\bibitem[{{Shields} {et~al}\mbox{.}(2009){Shields}, {Rosario}, {Smith},
  {Bonning}, {Salviander}, {Kalirai}, {Strickler}, {Ramirez-Ruiz}, {Dutton},
  {Treu}, \& {Marshall}}]{shields09}
{Shields} G.~A. {et~al.}, 2009, \apj, 707, 936

\bibitem[{{Sijacki}, {Springel} \& {Haehnelt}(2011){Sijacki}, {Springel}, \&
  {Haehnelt}}]{sijacki11}
{Sijacki} D., {Springel} V., {Haehnelt} M.~G., 2011, \mnras, 414, 3656

\bibitem[{{Smol{\v c}i{\'c}} {et~al}\mbox{.}(2014){Smol{\v c}i{\'c}},
  {Ciliegi}, {Jeli{\'c}}, {Bondi}, {Schinnerer}, {Carilli}, {Riechers},
  {Salvato}, {Brkovi{\'c}}, {Capak}, {Ilbert}, {Karim}, {McCracken}, \&
  {Scoville}}]{smolcic2014}
{Smol{\v c}i{\'c}} V. {et~al.}, 2014, \mnras, 443, 2590

\bibitem[{{Spergel} {et~al}\mbox{.}(2007){Spergel}, {Bean}, {Dor{\'e}},
  {Nolta}, {Bennett}, {Dunkley}, {Hinshaw}, {Jarosik}, {Komatsu}, {Page},
  {Peiris}, {Verde}, {Halpern}, {Hill}, {Kogut}, {Limon}, {Meyer}, {Odegard},
  {Tucker}, {Weiland}, {Wollack}, \& {Wright}}]{spergel2007}
{Spergel} D.~N. {et~al.}, 2007, \apjs, 170, 377

\bibitem[{{Steinhardt} {et~al}\mbox{.}(2012){Steinhardt}, {Schramm},
  {Silverman}, {Alexandroff}, {Capak}, {Civano}, {Elvis}, {Masters},
  {Mobasher}, {Pattarakijwanich}, \& {Strauss}}]{steinhardt12}
{Steinhardt} C.~L. {et~al.}, 2012, \apj, 759, 24

\bibitem[{{van der Laan} \& {Perola}(1969)}]{laan1969}
{van der Laan} H., {Perola} G.~C., 1969, \aap, 3, 468

\bibitem[{{Volonteri}, {Haardt} \& {Madau}(2003){Volonteri}, {Haardt}, \&
  {Madau}}]{volonteri03}
{Volonteri} M., {Haardt} F., {Madau} P., 2003, \apj, 582, 559

\bibitem[{{Wang} \& {Loeb}(2014)}]{wang2014}
{Wang} X., {Loeb} A., 2014, \mnras, 441, 809

\bibitem[{{Wright}(2006)}]{wright2006}
{Wright} E.~L., 2006, \pasp, 118, 1711

\bibitem[{{Wrobel}, {Comerford} \& {Middelberg}(2014){Wrobel}, {Comerford}, \&
  {Middelberg}}]{wrobel14}
{Wrobel} J.~M., {Comerford} J.~M., {Middelberg} E., 2014, \apj, 782, 116

\end{thebibliography}

\appendix

\bsp

\label{lastpage}

\end{document}